\newcommand{\bea}{\begin{eqnarray}}
\newcommand{\eea}{\end{eqnarray}}

\renewcommand{\a}{\alpha}

\newcommand{\sib}{Si$_\mathrm{B}$}
\newcommand{\on}{O$_\mathrm{N}$}
\newcommand{\ef}{E$_\mathrm{F}$}
\newcommand{\dd}{$d_\mathrm{{nn}}$}
\newcommand{\efm}{E$_{\mathrm{FM}}$}
\newcommand{\eam}{E$_{\mathrm{AFM}}$}
\newcommand{\ensp}{E$_{\mathrm{NSP}}$}
\newcommand{\meafm}{ME$_{\mathrm{AFM}}$}
\newcommand{\mefm}{ME$_{\mathrm{FM}}$}

\newcommand{\bt}[1]{{\bar t}}

\documentclass[twocolumn,showpacs,preprintnumbers,amsmath,amssymb]{revtex4}
\usepackage{graphicx}% Include figure files
\usepackage{dcolumn}% Align table columns on decimal point
\usepackage{bm}% bold math
\begin{document}

%\preprint{}

\title{Magnetism in 2D BN$_{1-x}$O$_x$ and B$_{1-x}$Si$_x$N: \\ polarized itinerant and local electrons}% Force line breaks with \\

\author{Ru-Fen Liu}
 \email{fmliu@phys.ncku.edu.tw;ru-fen.liu@crm2.uhp-nancy.fr}%Lines break automatically or can be forced with \\
% \altaffilation{Current address: CRM2, Insitut Jean Barriol,
%Nancy University and CNRS, 54506 Vand{\oe}uvre-l\`{e}s-Nancy, France}
\author{Ching Cheng}%
 \email{ccheng@mail.ncku.edu.tw}
\affiliation{%
Department of Physics, National Cheng Kung University,\\ 1 University Road, Tainan 70101, Taiwan }%

%\vskip1cm {\bf\Large \today \hskip5mm  File name apss$-$new2.tex}
%\vskip1cm

\begin{abstract}
We use density functional theory based first-principles methods to
study the magnetism in a 2D hexagonal BN sheet induced by the
different concentrations of oxygen and silicon atoms substituting
for nitrogen (\on) and boron (\sib) respectively. We demonstrate the
possible formation of three distinct phases based on the
magnetization energy calculated self-consistently for the
ferromagnetic (\mefm) and antiferromagnetic (\meafm) states, i.e.
the paramagnetic phase with \mefm=\meafm, the ferromagnetic phase
with \mefm$>$\meafm\ and finally the polarized itinerant electrons
with finite \mefm\ but zero \meafm. While the \on\ system was found
to exist in all three phases, no tendency towards the formation of
the polarized itinerant electrons was observed for the \sib\ system
though the existence of the other two phases was ascertained. The
different behavior of these two systems is associated with the
diverse features in the magnetization energy as a function of the
oxygen and silicon concentrations. Finally, the robustness of the
polarized itinerant electron phase is also discussed with respect to
the O substitute atom distributions and the applied strains to the
system.

\end{abstract}

% 75.75.+a Magnetic properties of nanostructures
% 61.46.-w Nanoscale materials
% 73.22.-f Electronic structure of nanoscale materials: clusters, nanoparticles, nanotubes, and nanocrystals Electronic structure of nanoscale materials: clusters, nanoparticles, nanotubes, and nanocrystals
% 75.70.Ak Magnetic properties of monolayers and thin films
% 61.72.Ji Point defects (vacancies, interstitials, color centers, etc.) and defect clusters
% 67.80.dj Defects, impurities, and diffusion
% 75.30.Hx Magnetic impurity interactions
% 71.10.Ca Electron gas, Fermi gas
% 71.10.Pm Fermions in reduced dimensions (condensed matter)
% 68.55.Ln Defects and impurities: doping, implantation, distribution, concentration, etc. (for diffusion of impurities, see 66.30.J?)
% 61.72.uj
% 61.72.-y Defects and impurities in crystals; microstructure

\maketitle

\section{Introduction}
The theoretical model of the two-dimensional (2D) homogeneous
electron gas (HEG), which considers the electrons moving in a 2D
uniform positive charge background, serves as an important model
system for studying the fundamental many-electron behavior in 2D
systems. These physical properties are closely connected to the
operations of conduction electrons in the layered semiconductor
devices as well as the recently synthesized 2D periodic
systems\cite{2dbn,Iijima09}. The most creditable method considered
presently for studying the 2D HEG is the quantum Monte
Carlo\cite{ceperley99,tanatar89,bachelet02,drummand08} (QMC) method.
The recent QMC studies found no region of stability for a
ferromagnetic fluid in the 2D HEG\cite{drummand08}. The phases were
found to transit directly from a polarized Wigner
crystal\cite{wigner} to the normal fluid (paramagnetic state).
However, it would be interesting to explore the parallel phase
diagram for real materials, i.e. systems of electrons moving in a
neutralizing background formed by a 2D lattice of nuclei, and study
the corresponding magnetic property as a function of the electron
density.

Inspired by the exciting experimental works on the discovery and
synthesization of hexagonal boron nitride (h-BN) in low-dimensional
structures\cite{Louie95,Iijima09,2dbn,ZGchen09} as well as the
possible formation of magnetism in these systems through creations
of defects\cite{RFL&CC,peralta08}, here we present Density
Functional Theory (DFT)\cite{Hohenber} based first-principles
studies on the magnetic properties of a 2D hexagonal BN sheet
induced by the oxygen and silicon atoms substituting for the N and B
atoms respectively (denoted as \on\ and \sib\ hereafter), i.e.
BN$_{1-x}$O$_x$ and B$_{1-x}$Si$_x$N. We demonstrate the possible
formation of polarized itinerant electrons in the \on\ system. The
stable phases are identified as, in increasing substitute atom
concentration $x$, the paramagnetic phase, the ferromagnetic phase
and finally the polarized itinerant electrons. In addition, the
\sib\ system which though supports the formation of ferromagnetic
phase displays no tendency to the stabilization of the polarized
itinerant electrons, i.e. resembling the QMC result for a 2D HEG.

\section{Computation Method}
The DFT calculations employed in the present study use the
generalized gradient approximation (GGA)\cite{GGA} for the
exchange-correlation energy functional and the projector
augmented-wave (PAW) method\cite{PAW,pawimp} for describing the
interaction between core and valence electrons as implemented in the
VASP\cite{vasp} code. To simulate the systems with the substitute
atom concentration $x$ ranging from $1/55$ to $1/4$, the supercells
with distance (denoted as \dd) of $\sim$17.5\AA, 13.3\AA, 10\AA,
7.5\AA\ and 5\AA\ between nearest-neighbor substitute atoms were
constructed. A vacuum distance larger than 12\AA\ was used in the
calculations to remove the interaction between layers. The numerical
convergence was accomplished by using k-points meshes of (7 7 1)
generated from Monkhost-Pack method\cite{Monk} for the supercell
consisting of $2\times2$ unit cells (and the similar density of
k-points meshes for the supercells of other sizes), and the
kinetic-energy cut-off of 400eV (with tests at 500 eV) for expanding
the single electron Kohn-Sham wavefunctions in plane-wave basis. The
atomic forces calculated by Hellmann-Feynman theorem\cite{force}
were relaxed to less than 0.02eV/\AA\ in all calculations.
Furthermore, the volume and shape were also allowed to relax for the
systems constructed by 2$\times$2 ($x=1/4$ and \dd$\sim$5\AA) and
3$\times$3 ($x=1/9$ and \dd$\sim$7.5\AA) supercells. We shall
emphasize that the relaxed structures for \on\ and \sib\ are found
to likely preserve its three-fold symmetry in the hexagonal lattice
at all studied substitution concentrations\cite{stability}. When
switching on the spin-polarized calculations, the atomic forces were
further relaxed.

\begin{figure}
\includegraphics{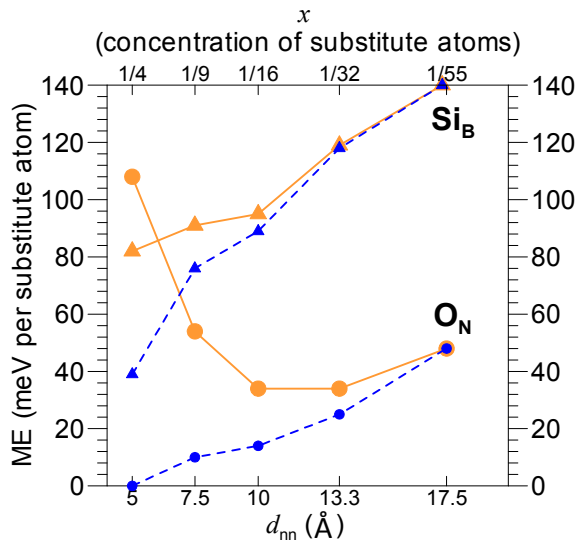}% Here is how to import EPS art
\caption{\label{emJ} The magnetization energy \mefm\ (solid lines)
and \meafm\ (dashed lines) per substitute atom with respect to \dd\
as well as the substitute atom concentration $x$. The circles and
triangles are for the systems of \on\ and \sib\ respectively.}
\end{figure}

\section{Results}
The calculated magnetizations for both \on\ and \sib\ stay as
1$\mu_B$ per substitute atom throughout the systems of fractional
$x$'s studied in this work\cite{graphene}. The magnetic moment found
in both systems can be attributed to the donated electrons
accommodating at the bottom of the conduction band due to the
presence of the substitute atoms which is confirmed by the
corresponding density of states (DOS) of these systems. The
magnetization energies (ME) for the ferromagnetic and
antiferromagnetic configurations are defined as the energy gains of
these configurations (\efm\ and \eam) with respect to those obtained
from the non-spin-polarized calculations (\ensp), i.e.
\mefm$=$E$_{\mathrm{NSP}}-$E$_{\mathrm{FM}}$ and
\meafm$=$E$_{\mathrm{NSP}}-$E$_{\mathrm{AFM}}$ respectively. (All
the magnetic related quantities discussed in this work are given in
the unit of per substitute atom.) Assuming that the interaction
among the local magnetic moments in these magnetic systems can be
approximately described by the nearest-neighbor Heisenberg
Hamiltonian, $H = -\sum_{i>j}JS_iS_j$ , then the interaction
strength as well as the magnetic coupling types, i.e. ferromagnetism
and anti-ferromagnetism, can be extracted from the exchange energy
$J=\frac{1}{4}($\mefm$-$\meafm$)$. Note that the initial magnetic
configurations in calculating \eam\ were set up with antiparallel
local moments of equal magnitude and then followed by the
spin-density relaxation in the self-consistent calculations. In
FIG.\ref{emJ} we present the calculated magnetization energy of \on\
and \sib\ as a function of the substitute atom concentration. At low
concentrations of $x\leq 1/55$ both systems are stabilized at
paramagnetic phase, i.e. $J=0$ but with finite \mefm=\meafm\ of
around 50meV for \on\ and 140meV for \sib. The presence of
substitute atoms in these concentrations therefore introduces local
moments into these systems yet no interactions among them, i.e.
isolated local moments. These moments were identified as mostly
$p_z$-like and their locality was manifested from the narrow
band-width bands splitting from the original conduction band of the
BN sheet. The spin density for \sib\ distributes mostly around the
Si atom sites, while that for \on\ is considerably more extensive,
i.e. up to the third nearest neighboring B atoms to the O atom.

As the substitute atom concentration is increased, both systems
develop into the ferromagnetic phase with finite $J$.  The values of
4$J$ can be as high as 42meV for \sib\ at $x=1/4$, which corresponds
to a Curie temperature estimated with the mean field approximation
($T_{CMF}$) of 80K and 44meV for \on\ at $x=1/9$, to a $T_{CMF}$ of
83K. Note that these ferromagnetic systems consist of no magnetic
elements like Fe, Co, Ni, or rare-earth elements.

The most particular feature takes place at $x=1/4$ of \on\ when a
finite \mefm\ is accompanied by a zero \meafm.  That is, the initial
antiparallel configurations in calculating E$_\mathrm{AFM}$ were
found to relax to the non-spin-polarized state. The zero \meafm\
together with a finite \mefm\ suggests the formation of polarized
itinerant electrons in \on, with an exchange split of 0.75eV in
bands.

\begin{widetext}
\begin{figure*}
\includegraphics{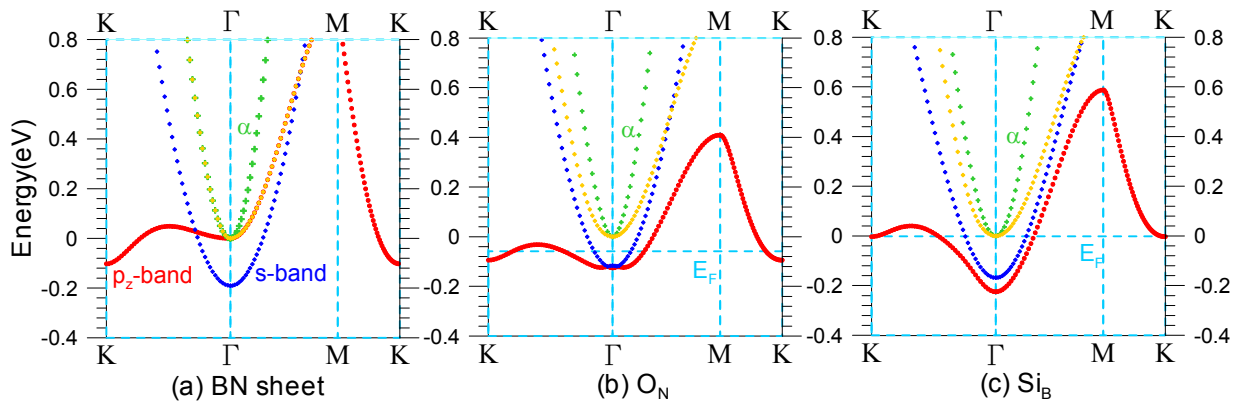}% Here is how to import EPS art
\caption{\label{band} Band structures for (a) BN sheet and (b) \on\
(c) \sib\ calculated at $x=1/4$. The energy zero is taken to be the
minimum of the band $\a$'s (green diamonds). }
\end{figure*}
\end{widetext}

The different characteristics of \on\ and \sib\ at $x=1/4$ are
implied in the distinct band structures generated by the
non-spin-polarized calculations. The band structure of hexagonal BN
sheet has been shown previously\cite{band} to display a
nearly-free-electron like characteristic around $\Gamma$ point of
the conduction band (CB). Since the most important contribution to
determine the magnetic properties in \on\ and \sib\ comes from the
electronic states near the Fermi level (\ef), we present in
FIG.\ref{band} the corresponding CBs for the BN sheet, and those for
\on\ and \sib\ at $x=1/4$. In the BN sheet, only $s$-electrons of N
atoms contribute to the free-electron like $s$-band (blue diamonds),
while the $p_z$-band (red dots) is mainly from the $p_z$-electrons
of B atoms. When the O or Si substitute atoms are introduced into
the system, the original $s$-band and $p_z$-band receive
contributions from both B and O atoms in \on\ but from Si atom alone
in \sib. Relative to the generally inert $\a$-band (whose energy at
$\Gamma$ point is taken as the energy reference zero in
FIG.\ref{band}), the effect of O and Si substitute atoms are
primarily the shift of $s$- and $p_z$-band and the change in
dispersion of the $p_z$-band. In \on\ the upward shifted $s$-band
and the downward shifted $p_z$-band results in the joint of the two
bands around the $\Gamma$ point. The joint along with a relatively
flatter $p_z$-band provide an excellent condition for high DOS at
\ef\ which further leads to the formation of polarized itinerant
electrons. In \sib, the main effect is the pressed downward
$p_z$-band to below the almost unaltered $s$-band with, contrary to
that in \on, a more dispersive $p_z$-band, i.e. a wider band width
for the occupied electrons in the CBs. In the previous GW band
structures calculations for BN sheet\cite{band} and h-BN\cite{hBN},
the effect of many-body correction to the bottom of CBs is mainly in
shifting the band energy rather than their curvatures. These results
would support our above conclusions as the join/disjoin of the $p_z$
and $s$-band as well as the curvature of the $p_z$-band dominate the
crucial physics of magnetism in \on\ and \sib.

The establishment of the polarized itinerant electrons in \on\ at
$x=1/4$ is further supported through the charge distribution for the
states in CBs just below \ef\ around $\Gamma$ point in the
non-spin-polarized calculation.  In \on, the charges distribute
mainly on the B atoms surrounding the O impurity (a rather extending
distribution) which is further joined by the charges distributing
above the more distant N atoms. Therefore the charges distribute as
connected charge sheet throughout the system for the isosurface of
density less than 0.324$e$\cite{isochg}. This result contrasts to
that in the \sib\ system whose charges distribute mostly around the
substitute Si atoms (a more local distribution) and the connected
isosurface occur only when the density is less than 0.213$e$.

The $x$-dependence ME's are also very different for \sib\ and \on.
In \sib, both the \mefm\ and \meafm\ decreases monotonically with
respect to increasing $x$ (decreasing \dd).  However, in \on, the
two $x$-dependence ME's diverge at values of $x$ larger than $1/16$,
i.e. the decreasing and approaching zero \meafm\ versus an
increasing \mefm.  We should emphasize here that the behavior of
\mefm\ for \on\ differs not only from that for \sib\ but also from
that for the systems carrying localized moments formed by other
non-magnetic substitute atoms in the BN sheet. Examples are the 2D
BN sheet with carbon atoms substituting for either B or N atoms, Si
atoms for N atoms, as well as the vacancies created by removing
either B or N atoms\cite{RFL&CC}. The exceptional behavior of \mefm\
together with the vanishing \meafm\ in \on\ provides a prominent
feature for this 2D system as developing into the polarized
itinerant electron phase.

\begin{figure}
\includegraphics{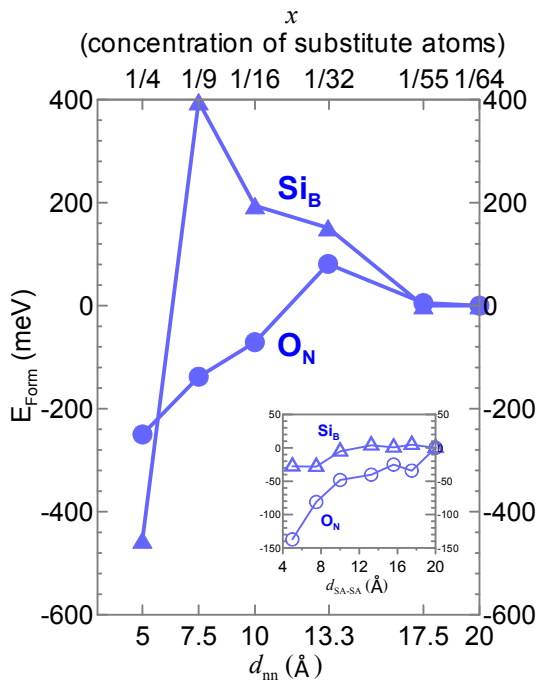}% Here is how to import EPS art
\caption{\label{eform} The $x$-dependence formation energy for \on\
and \sib, where the reference zero is chosen as the formation energy
at $x=1/64$, i.e. \dd=20\AA. }
\end{figure}

\section{Robustness of the polarized itinerant electrons in $O_N$}
In order to examine how the establishment of the polarized itinerant
electrons in \on\ at $x=1/4$ depends on the oxygen distributions,
different supercells consisting of O atoms with the same
nearest-neighbor distance \dd$\sim$5\AA\ but different numbers and
distributions are studied. They include a) a 6$\times$6 supercell
but with only eight instead of nine O substitute atoms, b) a
4$\times$4 supercell but with only three instead of four O atoms c)
a rectangular instead of a 4$\times$2 hexagonal supercell with two O
substitute atoms.  All these examined systems were found to be still
stabilized in the polarized itinerant electron phase with less than
4meV changes in \mefm. Regarding the substitute atom concentration
$x$, the highest $x$ of magnetic \on\ system we considered is $1/3$,
i.e. \dd=4.3\AA. At this concentration, the system remains
stabilized in the polarized itinerant electron phase. However, we
should emphasize that the role of N atoms in \on\ is essential as
magnetism would disappear if all N atoms are replaced by O atoms,
i.e. the BO sheet is in fact a normal metal. When mapping into the
2D HEG, this corresponds to the paramagnetic fluid at high electron
density.

FIG.\ref{eform} shows the $x$-dependence formation energy for \on\
and \sib, which is defined as \bea \mathrm{E}_{\mathrm{Form}}\equiv
\mathrm{E}_{\mathrm{O_N(Si_B)}}-
\mathrm{E}_{\mathrm{BNsheet}}+\mathrm{E}_\mathrm{N(B)}-\mathrm{E}_\mathrm{O(Si)},
\nonumber \eea where $\mathrm{E}_{\mathrm{O_N(Si_B)}}$ and
$\mathrm{E}_{\mathrm{BNsheet}}$ are the energies for the \on\ (\sib)
system (ferromagnetically polarized ones) and the BN sheet
calculated by using exactly the same supercell to eliminate
numerical errors, E$_\mathrm{N}$ (E$_\mathrm{B}$) and E$_\mathrm{O}$
(E$_\mathrm{Si}$) are the energies for the removed N (B) and the
added O (Si) atoms respectively\cite{eform}. The formation energy
for the system of $x=1/64$, i.e. the system with no interaction
among the substitute atoms, is taken as the reference zero in the
figure. Unlike the repulsive interactions in \sib\ at $x\leq 1/9$,
the O atoms introduced for substituting N atoms in the BN sheet
exhibit attractive interaction once $x$ becomes larger than $1/32$.
Besides, by employing a very large supercell (11$\times$10) with
only two substitute atoms (SA) created at the distance
$d_{\mathrm{SA-SA}}$ in the cell, it is shown (see the insert in
FIG.\ref{eform}) that for both systems the substitute atoms tend to
aggregate to form the system of $x=1/4$.

Considering the possible experimental fabrication of this 2D system
supported on a substrate, our study show that the system with
polarized itinerant electrons can sustain an external applied
tensile and compressive strain of up to 0.1, i.e. a change of 10\%
in lattice constant. Similar study for \sib, on the contrary, leads
to a non-magnetic state when the system is under compressive strain.

\section{Conclusion}
In summary, we have explored the possible magnetic phases in the
\on\ and \sib\ systems using first-principles methods.  The
polarized itinerant electron phase was found in the
BN$_{3/4}$O$_{1/4}$ system while in the \sib\ system only the
paramagnetic phase at low silicon concentrations and ferromagnetic
phase at high silicon concentrations were identified. The different
magnetic properties in \on\ and \sib\ were discussed through the
$x$-dependence magnetization energy, the band structures, and the
charge-density distributions. The robustness of the formation of the
polarized itinerant electron phase was also discussed with respect
to the O substitute atom distributions and the applied strain to the
system.

\acknowledgements This work was supported by the National Science
Council of Taiwan. Part of the computer resources are provided by
the NCHC (National Center of High-performance Computing). We also
thank the support of NCTS (National Center of Theoretical Sciences)
through the CMR (Computational Material Research) focus group.


\begin{thebibliography}{99}
\bibitem{2dbn}
K. S. Novoselov, D. Jiang, F. Schedin, T. J. Booth, V. V.
Khotkevich, S. V. Morozov, and A. K. Geim, Proc. Natl Acad. Sci. USA
{\bf 102}, 10451 (2005).
\bibitem{Iijima09}
Chuanhong Jin, Fang Lin, Kazu Suenaga, and Sumio Iijima, Phys. Rev.
Lett. {\bf 102}, 195505 (2009).
\bibitem{ceperley99}
D. M. Ceperley, Nature, {\bf 397}, 386 (1999).
\bibitem{tanatar89}
B. Tanatar and D. M. Ceperley, Phys. Rev. B, {\bf 39}, 5005 (1989).
\bibitem{bachelet02}
C. Attaccalite, S. Moroni, P. Gori-Giorgi, and G. B. Bachelet, Phys.
Rev. Lett., {\bf 88}, 256601 (2002).
\bibitem{drummand08}
N. D. Drummond and R. J. Needs, Phys. Rev. Lett. {\bf 102}, 126402
(2009).
\bibitem{wigner}
E. P. Wigner, Phys. Rev. {\bf 46}, 1002 (1934).
\bibitem{Louie95}
N.G. Chopra, R.J. Luyken, K. Cherrey, V.H. Crespi, M.L. Cohen, S.G.
Louie, and A. Zettl, Science {\bf 269}, 966 (1995).
\bibitem{ZGchen09}
Zhi-Gang Chen, Jin Zou, Gang Liu, Feng Li, Yong Wang, Lianzhou Wang,
Xiao-Li Yuan, Takashi Sekiguchi, Hui-Ming Cheng and Gao Qing Lu, ACS
Nano {\bf 2}, 2183 (2008).
\bibitem{RFL&CC}
Ru-Fen Liu and Ching Cheng, Phys. Rev. B {\bf 76}, 014405 (2007).
\bibitem{peralta08}
Veronica Barone and Juan E. Peralta, Nano Lett. {\bf 8}, 2210
(2008).
\bibitem{Hohenber}
P. Hohenberg and W. Kohn, Phys. Rev., {\bf 136}, B864 (1964); W.
Kohn and L. J. Sham, Phys. Rev., {\bf 140}, A1133 (1965).
\bibitem{GGA}
J. P. Perdew in '\textit{Electronic Structure of Solids '91}, edited
by P. Ziesche and H. Eschrig (Akademie-Verlag, Berlin, 1991); J. P.
Perdew et al., Phys. Rev. B {\bf 46}, 6671 (1992).
\bibitem{PAW}
P.E. Bl\"{o}chl, Phys. Rev. B {\bf 50}, 17953 (1994).
\bibitem{pawimp}
G. Kresse and D. Joubert, Phys. Rev. B {\bf 59}, 1758 (1999).
\bibitem{vasp}
Vienna \textit{ab initio} Simulation Package, G. Kresse and J.
Hafner, Phys. Rev. B {\bf 47}, 558 (1993); {\bf 49}, 14251 (1994);
G. Kresse and J. Furthmuller, Comput. Mater. Sci. {\bf 6}, 15
(1996); Phys. Rev. B {\bf 54}, 11169 (1996).
\bibitem{Monk}
H. J. Monkhorst and J. D. Pack, Phys. Rev. B {\bf 13}, 5188 (1976).
\bibitem{force}
H. Hellmann, \textit{Einf\"{u}hrung in die Quantenchemie} (Deuticke,
Leipzig, 1937), pp.61 and 285; R. P. Feynman, Phys. Rev. {\bf 56},
340 (1939).
\bibitem{stability}
A class of the doped systems involving only 2$s$- and 2$p$-
electrons in the host 2D BN sheet has been studied
previously\cite{RFL&CC}, in which two types of magnetic moments are
identified, i.e. the planer one formed by $sp2$-electrons and the
perpendicular one formed by $p_z$-electrons. The structure likely
undergoes distortions for the systems with planer magnetic moments,
besides we further found that their moments tend to vanish at
$x=1/4$; while that for the systems with perpendicular magnetic
moments likely stays in the three-fold symmetry.
\bibitem{graphene}
Contrary to the invariable magnetic moments of 1$\mu_B$ by the
substituted O and Si atoms in a BN sheet, the magnetic moments by
substituting B and N atoms in graphene disappear at low substitution
concentration\cite{signh09}. This could be attributed to the unusual
linear dispersion $\pi$ and $\pi$* bands joined at the K symmetry
point in reciprocal space in graphene.
\bibitem{signh09}
Ranber Singh and Peter Kroll,  J. Phys.: Condens. Matter, {\bf 21},
196002 (2009).
\bibitem{band}
X. Blase, Angel Rubio, Steven G. Louie, and Marvin L. Cohen, Phys.
Rev. B {\bf 51}, 6868 (1995).
\bibitem{hBN}
B. Arnaud, S. Leb\`{e}gue, P. Rabiller, and M. Alouani, Phys. Rev.
Lett., {\bf 96}, 026402 (2006).
\bibitem{isochg}
The distance between two layers in the bulk hexagonal BN, i.e.
3.3\AA, were used to determine the isocharge density.
\bibitem{eform}
There are probably more appropriate choices for E$_\mathrm{N(B)}$
and E$_\mathrm{O(Si)}$, e.g. half of the energy of O$_2$ for
E$_\mathrm{O}$. However, as we are discussing the formation energy
relative to that of the systems with $x=1/64$, the systems with
those terms are cancelled out and play no role here.


\end{thebibliography}
\end{document}